\documentclass[9pt,twocolumn,twoside]{pnas-new}

\templatetype{pnasresearcharticle} 

\title{A unified energy optimality criterion predicts human navigation paths and speeds}

\author[a,b,1]{Geoffrey L. Brown}
\author[a,c,1]{Nidhi Seethapathi}
\author[a,2]{Manoj Srinivasan}

\affil[1]{\textbf{Equal contribution}}
\affil[a]{Mechanical and Aerospace Engineering, the Ohio State University, Columbus, OH 43210}
\affil[b]{Feinberg School of Medicine, Northwestern University, Chicago, IL 60611}
\affil[c]{Department of Bioengineering, University of Pennsylvania, Philadelphia, PA 19104}

\leadauthor{Brown, Seethapathi, Srinivasan}

\significancestatement{Why do humans move the way they do? Here, we obtain a physiologically-based theory of the speeds and paths with which humans navigate their environment. For the first time, we measure the metabolic energy cost of walking with turning and show that minimizing this cost explains diverse phenomena involving navigating around obstacles, walking in complex paths, and turning. We explain why humans slow down while turning, avoid sharp turns, do not always use the shortest path, and other naturalistic locomotor phenomena.}

\authorcontributions{Conceptualization, GB, NS, and MS; Methodology, GB, NS, and MS; Software and Formal Analysis, GB, NS, and MS; Writing -- Original Draft, GB and MS; Writing -- Review \& Editing, NS and MS; Supervision, MS; Funding Acquisition, MS.}
\authordeclaration{The authors declare that they have no competing interests.}
\equalauthors{\textsuperscript{1}G.B. contributed equally to this work with N.S..}
\correspondingauthor{\textsuperscript{2}To whom correspondence should be addressed. E-mail: srinivasan.88\@osu.edu}


\begin{abstract}
Navigating our physical environment requires changing directions and turning. Despite its ecological importance, we do not have a unified theoretical account of non-straight-line human movement. Here, we present a unified optimality criterion that predicts disparate non-straight-line walking phenomena, with straight-line walking as a special case. We first characterized the metabolic cost of turning, deriving the cost landscape as a function of turning radius and rate. We then generalized this cost landscape to arbitrarily complex trajectories, allowing the velocity direction to deviate from body orientation (holonomic walking). We used this generalized optimality criterion to mathematically predict movement patterns in multiple contexts of varying complexity: walking on prescribed paths, turning in place, navigating an angled corridor, navigating freely with end-point constraints, walking through doors, and navigating around obstacles. In these tasks, humans moved at speeds and paths predicted by our optimality criterion, slowing down to turn and never using sharp turns. We show that the shortest path between two points is, counterintuitively, often not energy optimal, and indeed, humans do not use the shortest path in such cases.  Thus, we have obtained a unified theoretical account that predicts human walking paths and speeds in diverse contexts. Our model focuses on walking in healthy adults; future work could generalize this model to other human populations, other animals, and other locomotor tasks.
\end{abstract} 

\dates{This manuscript was compiled on \today}
\doi{\url{www.pnas.org/cgi/doi/10.1073/pnas.XXXXXXXXXX}}

\begin{document}

\maketitle
\thispagestyle{firststyle}
\ifthenelse{\boolean{shortarticle}}{\ifthenelse{\boolean{singlecolumn}}{\abscontentformatted}{\abscontent}}{}

\dropcap{R}eal-world human navigation through our physical environment requires changing direction and turning, maneuvering around obstacles, and moving along complex paths. In one previous study that tracked indoor locomotion over many days, 35-45\% of steps required turns \cite{glaister2007video}. Despite the importance of turning in ecological settings, we do not have a coherent theoretical account of the paths and speeds observed in such locomotion. Human subject experiments \cite{Rals58,Don01,LongSrinivasan2013,seethapathi2015metabolic,handford2014sideways,selinger2019humans} and mathematical models \cite{Min97,Kuo01,Sri06,Bog10,Sri11,Miller12,seethapathi2015metabolic,JoshiSrinivasan2015BridgeProcA,falisse2019rapid} have suggested that energy optimality explains many aspects of straight line locomotion, at least approximately. However, we do not know if such energy optimality generalizes to walking while navigating a more complex environment. Here, we obtain a better understanding of locomotion with turning, first quantifying its increased energetic demands and then showing that accounting for these increased energetic demands correctly predicts human navigation paths and speeds in a variety of naturalistic locomotor contexts.

Over the years, researchers have measured human locomotion in a few contexts requiring changing direction, for instance, navigating through angled corridors \cite{dias2014pedestrian}, moving from point-to-point while having to turn \cite{mombaur2010human}, walking through doors,  and avoiding obstacles \cite{Pham07,Are08}. However, previous models aimed at explaining such data used minimization principles that were not physiologically-based, required fitting model parameters to behavioral data, were generally fit to only one experiment, and did not generalize to multiple experiments  \cite{Pham07,Are08,mombaur2010human}. As we argue later on, these models could not simultaneously explain the paths and speeds observed in curvilinear walking, often predicting zero speeds for simple  tasks. Here, we provide a theoretical account that does not have these limitations and is broadly predictive.

We perform the first human subject experiments to quantify the metabolic cost of humans walking with turning, measuring how the metabolic cost increases with walking speed and the turning rate. We generalize this empirically-derived metabolic cost landscape to walking along arbitrary paths and then use an optimization-based framework to make a number of behavioral predictions about humans walking in tasks of different complexity. We compared our predictions to five different experiments, each containing a qualitatively different walking or turning task. These five experiments consist of two new behavioral experiments we performed here and data from three prior studies. Specifically, we predict that humans would walk slower when turning in smaller circles, which we compare with our own behavioral experiments, correctly predicting the lowered walking speeds. 
We show that the speed at which humans turn in place is approximately predicted by minimizing the cost of turning, again comparing with our own experiments. Finally, minimizing the same metabolic model, we predict more complex walking behavior involving navigation observed in three previous studies: walking freely from point to point \cite{mombaur2010human}, walking through doors and avoiding obstacles \cite{Pham07,Are08}, and walking and turning along an angled corridor \cite{dias2014pedestrian}. We show that energy optimality explains many qualitative and quantitative features of human walking including not taking sharp turns, path shapes adopted while walking with turning, and speed reductions during turns, as observed in these prior experimental studies \cite{Pham07,Are08,mombaur2010human,dias2014pedestrian}, but heretofore not predicted by a single theoretical model. Both our metabolic models and behavioral predictions focus on walking in healthy adults and we outline the benefits of generalizing our approach to other populations and other locomotor tasks.

\section*{Results}
\begin{figure*}[ht!]
\centering
\includegraphics{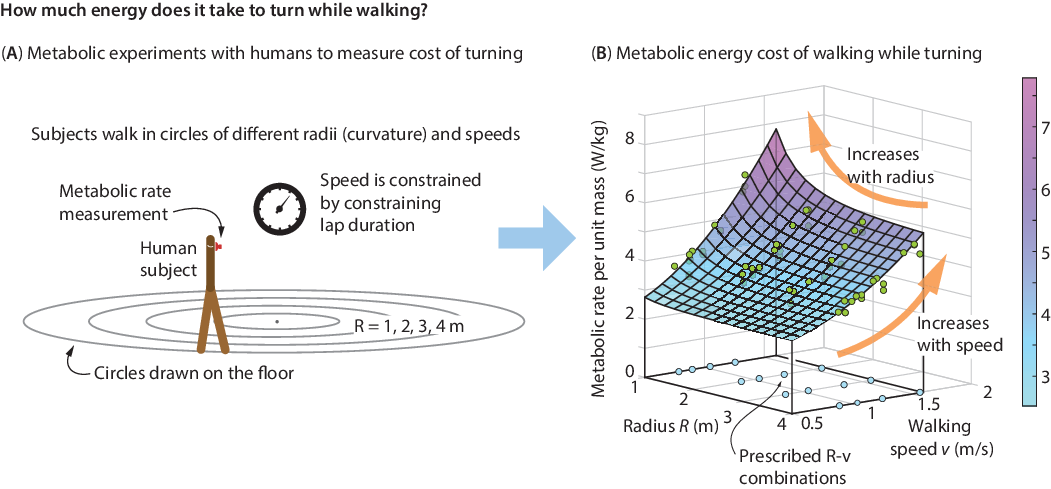}  
\caption{\textbf{Energy cost landscape of turning from humans walking in circles.} (\textbf{A}) We estimated the metabolic rate as a function of walking speed and turning radius by having subjects walk in circles of a few different radii and at a few different speeds at each radius. Speeds were constrained by having them complete laps at prescribed durations. Metabolic rate was estimated using respiratory gas analysis. (\textbf{B}) The metabolic rate data per unit body mass  ($\dot{E}$) is higher for higher speeds and lower radii. The prescribed speeds and radii $(v,R)$ at which the data was collected is shown as blue dots on the horizontal plane ($\dot{E}=0$ plane); raw metabolic data points from four representative subjects are shown as green dots (see \textit{SI} for all data). The wireframe surface shows the best-fit model $\dot{E} = \alpha_0 + \alpha_1v^2+\alpha_2 \omega^2$, capturing the nonlinear increase with both linear velocity $v$ and angular velocity $\omega$; the model captures 88.1\% of the data variance. The four corners of the cost surface has been connected to the horizontal plane to enhance the 3D visualization; a contour plot version of this surface is Supplementary Information Figure S1.}
\label{fig:MetExpt}
\end{figure*}

\paragraph{Turning increases metabolic cost substantially.} We measured the metabolic energy expenditure of seventeen human subjects while walking with turning. To measure the cost of turning, we instructed the subjects to walk in circles of different radii and at different tangential speeds along the circle. We provided feedback to them ensure that they were able to perform the required task (see Figure \ref{fig:MetExpt}A and the Methods section). Each subject performed at least 16 trials of 4 radii and at least 4 speeds. We measured metabolic energy expenditure using indirect calorimetry, that is, by tracking respiratory oxygen and carbon dioxide flux \cite{Brock87}. Figure \ref{fig:MetExpt}B shows the resulting mass-normalized metabolic rate, that is, metabolic energy per unit time per unit subject mass $\dot{E}$, as a function of prescribed speed $v$ and path radius $R$. These measurements show that for a given prescribed walking speed, the metabolic energy expenditure was higher for smaller radii, or equivalently, for higher path curvature (Figure \ref{fig:MetExpt}). For instance, directly comparing the measured metabolic rates at equal prescribed speeds but different radii, we find that walking at 1\,m radius was more expensive than walking at 4\,m radius by 0.59 W/kg on average ($p < 4 \times 10^{-4}$ using a right-tailed t-test), 1\,m radius was more expensive than 2\,m radius by 0.46 W/kg on average ($p  = 9 \times 10^{-4}$), and 2\,m radius was more expensive than 4\,m radius by 0.14 W/kg on average ($p < 10^{-6}$). These differences constitute large energy penalties relative to the resting metabolic rate: 40\%, 20\%, and 10\% respectively, corresponding to Cohen's $d$ effect sizes of 0.64, 0.62, and 0.15. Because these differences are computed at matched speeds (albeit prescribed speeds), the differences are significant for both metabolic cost per unit time and per unit distance around the circle. 

\paragraph{Metabolic rate increases nonlinearly with linear and angular speeds.}  The total metabolic rate was described well by the following quadratic function of the linear speed $v$ (tangent to the walking path) and the angular speed $\omega = v/R$ as follows (Figure \ref{fig:MetExpt}B): 
\begin{equation} 
\dot{E} = \alpha_0 +   \alpha_1 v^2 + \alpha_2 \frac{v^2}{R^2} =  \alpha_0 +   \alpha_1 v^2 + \alpha_2 \omega^2, \label{eq:MetRateFit} 
\end{equation}
 with $\alpha_0 = 2.204 \pm 0.079$ W/kg, $\alpha_1 = 1.213 \pm 0.054$ W/kg/(ms$^{-1})^2$, $\alpha_2 = 0.966 \pm  0.061$ W/kg/(rad.s$^{-1})^2$, giving the metabolic rate $\dot{E}$ in W/kg (normalized by body mass), where $v$ is in ms$^{-1}$, $R$ is in meters, and  $\omega$ is in rad.s$^{-1}$. The $p$ values for the three coefficients obeyed $p < 10^{-30}$, compared to a null constant model of zero coefficients. Equation \ref{eq:MetRateFit} explains 88.1\% of the empirical metabolic rate variance over all subjects (this percentage is the statistical $R$-squared value, the coefficient of determination, not to be confused with radius-squared). 
 
The form for the metabolic rate expression in equation \ref{eq:MetRateFit} was chosen in analogy to classic work on straight line walking \cite{Rals58,bobbert1960energy}, which found that the metabolic rate is close to linear in $v^2$, that is, $\dot{E} \sim \alpha_0 + \alpha_1 v^2$. In a post hoc analysis, we compared the 88.1\% variance explained by simple quadratic expression in equation  \ref{eq:MetRateFit} to more general quadratic expressions for $\dot{E}$ (with additional $v, |\omega|$ and $v|\omega|$ terms); such more general quadratic expressions increased the explained variance by less than 0.7\%. See Supplementary Information Table S1. We use the absolute value $|\omega|$ as we did not distinguish between leftward and rightward turns, ignoring any left-right gait asymmetries.  Among such more general quadratic models, the model in equation \ref{eq:MetRateFit} has the lowest Bayesian Information Criterion, a common model selection criterion that promotes model parsimony while maintaining a good fit to data  \cite{neath2012bayesian}. Similarly, allowing exponents other than 2 in the model expression, considering $\dot{E} \sim \alpha_0 + \alpha_1 v^\gamma + \alpha_2 \omega^\delta$, improves the explained variance by less than 0.4\% while adding two more model parameters and so we did not consider it further.
 
\begin{figure*}[ht!]
\centering
\includegraphics{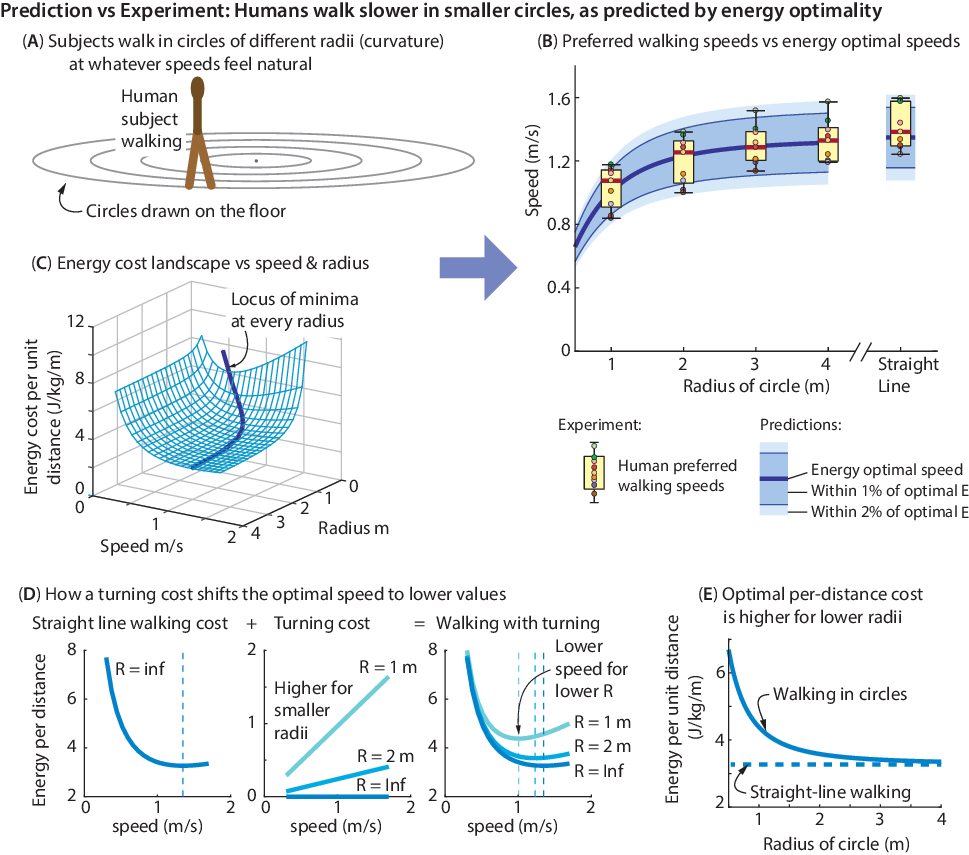}  
\caption{\textbf{Prediction vs behavior: Preferred walking in circles.} (\textbf{A})  To test behavioral predictions of energy optimal walking, we asked subjects to walk on circles of different radii at whatever speeds they found natural. (\textbf{B})  Human preferred walking speeds and model-predicted optimal walking speeds. Humans walk slower for smaller radii, as also predicted by energy optimality. The yellow box-plot shows human preferred walking speed along with individual data points (box indicates 25$^{\mathrm{th}}$, median, and 75$^{\mathrm{th}}$ percentile and whiskers indicate the range). The solid dark blue line is the optimal tangential speed $v_\mathrm{opt}$ for every radius. Also shown are two bands denoting speeds for which the metabolic cost per distance is within 1\% (lighter blue) and 2\% (darker blue) of the optimum cost. Most humans seem to be within 2\% of their energy optima. Sensitivity of these predictions to uncertainty in the metabolic model coefficients is shown in Supplementary Figure S2. (\textbf{C})  Metabolic cost per unit distance per unit mass. Minimizing this function at each radius produces model predictions in panel-b, identical to the dark blue line panel-c. (\textbf{D})  The turning cost per unit distance is linear in velocity ($\alpha_3 v/R$) and modifies the walking cost in a manner that the optimal speed is lower for smaller radii or higher curvatures. (\textbf{E})  The optimal metabolic cost per unit distance as a function of the radius. }
    \label{fig:MetCostPerDistannceEtc}
\end{figure*}

\paragraph{Straight-line walking is a less expensive special case.} Setting angular speed $\omega$ to zero or radius $R$ to infinity in equation~\ref{eq:MetRateFit} gives $\dot{E}$ for straight line walking: $\alpha_0 +  \alpha_1 v^2$.   Thus, as noted, the quadratic expression (Eq. \ref{eq:MetRateFit}) generalizes the classic quadratic expression for the straight-line walking metabolic rate, namely, $\alpha_0+\alpha_1 v^2$ \cite{Rals58}. Previous studies of overground or treadmill straight line walking \cite{Rals58,LongSrinivasan2013} have estimated $\alpha_0 \approx 2-2.5$ W/kg and $\alpha_1 \approx 0.9-1.4$ W/(kg.ms$^{-1}$), and our estimates are squarely in this same range. Because the coefficient $\alpha_2 > 0$ with $p<10^{-4}$, the model (Eq. \ref{eq:MetRateFit}) confirms that  the estimated metabolic rate is higher for lower radii $R$ for a given tangential speed $v$. This radius dependence implies, for instance, that at speed $v = 1.5$ m/s, reducing the radius $R$ from infinity to 1 m induces an additional cost ($\alpha_2 v^2/R^2$) of about 43\% of the total straight-line walking metabolic rate ($\alpha_0 +   \alpha_1 v^2$). This turning cost is about 60\% of the net straight-line walking metabolic rate ($\alpha_0 +   \alpha_1 v^2-\dot{E}_\mathrm{rest}$), that is, over and above the resting metabolic rate $\dot{E}_\mathrm{rest}$. 

As a brief aside, we note that the coefficient $\alpha_0$ has the interpretation of the metabolic rate of walking steadily at very low speeds ($v \approx 0$), which is substantially higher than, and should not be conflated with, the resting metabolic rate $\dot{E}_\mathrm{rest}$ (here, around $1.4$ W/kg). This difference between $\alpha_0$ and $\dot{E}_\mathrm{rest}$ is similar to differences found by previous studies between the cost of standing still and very slow walking with stepping \cite{Rals58}.

\paragraph{An outline: Energy-optimal behavioral predictions versus experiment.} In the rest of this section, we alternate between making behavioral predictions from energy optimization and comparing these predictions with experiment. First, we consider simple tasks such as walking in circles, turning in place, and walking in straight lines: tasks in which the movement path is fixed. For these tasks, we directly use the metabolic  model in equation \ref{eq:MetRateFit}. Then, we consider more complex tasks that require the movement paths to be selected by the subject: for these tasks, we first generalize the metabolic energy model in equation \ref{eq:MetRateFit} by incorporating other prior metabolic data, thereby accommodating more general locomotion and arbitrary paths.

\paragraph{Prediction: Optimal speeds are lower for smaller circles.}
When walking a long-enough distance in a straight line in the absence of time constraints, humans usually walk close to the speed that minimizes the total metabolic cost per unit distance $E^\prime = \dot{E}/v$ \cite{Rals58,Sri09,LongSrinivasan2013,seethapathi2015metabolic}. This energy optimal speed is sometimes called the maximum range speed \cite{Sri09} as it also maximizes the straight-line distance traveled with a fixed energy budget. Analogously, for walking in circles (Figure \ref{fig:MetCostPerDistannceEtc}A), we hypothesize that humans will use speeds that minimize the total cost per unit distance $E^\prime = \dot{E}/v = \alpha_0/v + (\alpha_1+\alpha_2/R^2) v$. This cost per unit distance is minimized when the slope $\partial E^\prime/\partial v = 0$, that is, at the speed $v_{\mathrm{opt}} = \sqrt{\alpha_0/ (\alpha_1 + \alpha_2/R^2)}$ (see Figure \ref{fig:MetCostPerDistannceEtc}B). This optimal speed $v_{\mathrm{opt}}$ is lower for lower radius $R$, thus predicting that humans would prefer to walk slower in smaller circles (Figure \ref{fig:MetCostPerDistannceEtc}B). Figures \ref{fig:MetCostPerDistannceEtc}C-D provide intuition for how the turning cost lowers the optimal speed for circle-walking. The quantity minimized here, namely the mass-normalized cost per unit distance $E^\prime$, is a scaled version of the cost of transport \cite{Sri06,Sri09}, a non-dimensional quantity given by $E^\prime/g$. 

\paragraph{Experiment: Human preferred speeds are lower in smaller circles as predicted by energy optimality.}  We asked people to walk naturally on circular paths of four different radii and in a straight line (Figure \ref{fig:MetCostPerDistannceEtc}A). As predicted by energy optimality, we found that humans preferred lower speeds for smaller circles (Figure \ref{fig:MetCostPerDistannceEtc}B; \cite{orendurff2006kinematics}). The median preferred speed across subjects is well predicted by the energy optimal speed at every radius, and essentially all of the preferred speeds at any radius are within 2\% of optimal energy cost (Figure \ref{fig:MetCostPerDistannceEtc}B).  In this walking-in-circles experiment, humans are clearly able to walk at faster or slower speeds than optimal at each radius, as demonstrated in our earlier metabolic trials (Figure \ref{fig:MetExpt}). For the special case of straight-line walking, the optimal speed is given by $v_{\mathrm{opt}} = \sqrt{\alpha_0/ \alpha_1 } = 1.35$ m/s, which agrees with typical human preferred walking speeds in previous studies \cite{Sri09,bohannon2011normal,seethapathi2015metabolic} as well as the trials here. Further, for every single subject and every single trial, the data was such that their average preferred speed for a larger radius circle was higher than that of a smaller circle ($v_\mathrm{pref, 4m} > v_\mathrm{pref, 3m} > v_\mathrm{pref, 2m} > v_\mathrm{pref, 1m}$).

\begin{figure*}[ht!]
\begin{center}
	\includegraphics{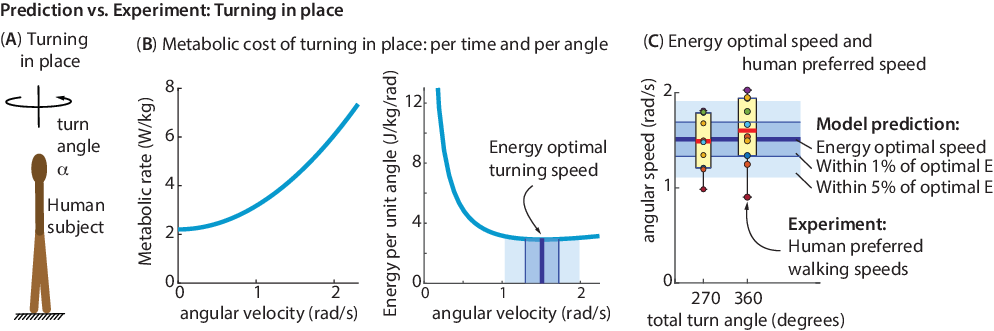}
\caption{\textbf{Prediction vs behavior: Turning in place.} (\textbf{A})  To further test behavioral predictions of energy optimality, we asked subjects to turn by a given angle $\alpha$, starting and ending at rest. (\textbf{B})  Metabolic rate and the cost per unit turning angle, obtained by extrapolating the model to turning-in-place. The blue lines and bands shown denote optimal turning speeds and the set of speeds within 1\% or 5\% of optimal energy cost. (\textbf{C})  Human preferred turning speeds (yellow box plot and individual data points) largely overlap with the turning speeds within 5\% of the optimal cost. Sensitivity of these predictions to uncertainty in the metabolic model coefficients is shown in Supplementary Figure S2.}
\label{fig:TurningInPlace}
\end{center}
\end{figure*}

\paragraph{Corollary: A straight line path is optimal if there are no other constraints or obstacles.} Conventional wisdom dictates that (in the absence of other constraints), a straight line path would be energy optimal to travel a given distance between two points A and B.  This conventional wisdom relies on implicit assumptions about the metabolic energy landscape unavailable before our measurements. Assume that the distance to be traveled is long-enough \cite{seethapathi2015metabolic} so that we minimize the metabolic cost per unit distance: $E^\prime = \dot{E}/v = \alpha_0/v + (\alpha_1+\alpha_2/R^2) v$, ignoring any small initial or final transient costs. Then, at any speed, this cost is minimized when the turning radius $R$ goes to infinity.  This result is reflected in Figure \ref{fig:MetCostPerDistannceEtc}E, which shows that the minimum cost per unit distance at any given turning radius, and the minimum is achieved when radius $R$ goes to infinity. Because the straight line both minimizes the distance traveled and the cost per unit distance, the straight line minimizes the total cost for a given distance. Such optimality of straight line walking will be true for any metabolic rate model that has positive cost penalty for turning, not necessarily the specific functional form we have considered here. However, as noted earlier, we considered functional forms for the metabolic cost that allowed turning to result in energy reduction e.g., negative $\alpha_2$ or a linear term in $|\omega|$ with a negative coefficient. Such metabolic cost functions were not justified by our metabolic data (Supplementary Table S1). Of course, the optimality of the straight line path is not generally true in the presence of obstacles or constraints such as those on initial and final body orientation, as considered later in this section. 

\paragraph{Prediction: Minimizing energy cost of turning-in-place predicts an optimal turning rate.} Humans often need to turn in place while standing, to re-orient their body --- to face a new direction. Turning in place or spinning in place (Figure \ref{fig:TurningInPlace}A) is a special case of walking in circles with radius $R \rightarrow 0$ and speed $v \rightarrow 0$, while the angular velocity $\omega = v/R$ remains non-zero. Extrapolating using these limits, we obtain the metabolic rate of turning-in-place to be: $\dot{E} = \alpha_0 + \alpha_2 \omega^2$. The metabolic cost of turning in place per unit angle (analogous to metabolic cost per unit distance) is $\dot{E}/\omega = \alpha_0/\omega + \alpha_2 \omega$. This cost per unit angle is optimized by steady optimal turning speed $\omega_{\mathrm{opt}} = \sqrt{\alpha_0/\alpha_2}$ =  1.46 rad/s = 83.6 degrees/s (Figure \ref{fig:TurningInPlace}B-C).

\paragraph{Experiment: Humans turn in place at close to the energy optimal turning rate.}  We performed behavioral experiments in which humans turned in place by a fixed turn angle $\alpha$ (Figure \ref{fig:TurningInPlace}A), starting and ending at rest. The average human turning speeds for large turns of 270 degrees and 360 degrees largely overlap with each other and almost entirely overlap with the set of steady turning speeds that are within 5\% of the optimal turning cost  (Figure \ref{fig:TurningInPlace}C). 

\begin{figure*}[ht!]
\begin{center}
	\includegraphics{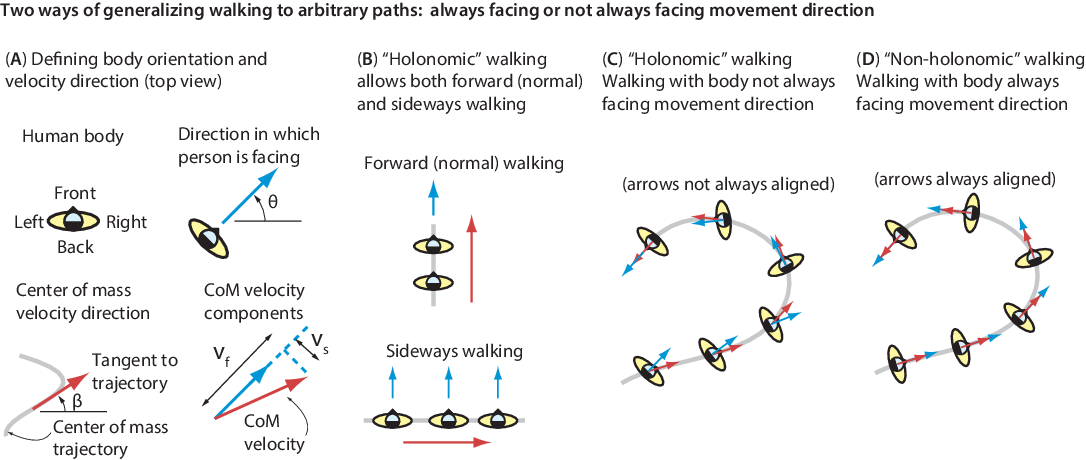}
\caption{\textbf{Two kinds of walking: face where you are going or not.} (\textbf{A})  Visual representation of the body from a top view, introducing different notations for the direction in which the body center of mass is moving (red arrow) and the body orientation (blue arrow). The components of the center of mass velocity are shown to be $v_f$ and $v_s$ in the anterior-posterior (forward) and medio-lateral (sideways) directions respectively. (\textbf{B})  While forward walking has body orientation aligned with movement direction, it is possible to walk sideways, so that the body orientation is perpendicular to the movement direction.  (\textbf{C})  Walking in a manner that that the body orientation need not be aligned with the movement direction (``holonomic''); that is, the blue and red arrows need not be aligned. (\textbf{D})  Walking in a manner that the body orientation is the same as the movement direction (``non-holonomic''); that is, the blue and red arrows are aligned. }
\label{fig:HolonomicExpository}
\end{center}
\end{figure*}

\paragraph{Two ways of generalizing to complex paths: face the movement direction or not.} We now generalize the metabolic cost of walking in circles to walking on arbitrary paths, but first, we discuss what assumptions such generalizations make. When walking, we can conceptually distinguish between the  direction in which our body moves (velocity direction, angle $\beta$, Figure \ref{fig:HolonomicExpository}A) and the direction in which the body faces (body torso orientation, angle $\theta$). For instance, in normal straight-line walking, these two directions are aligned ($\theta = \beta$), but in ``sideways walking'' \cite{handford2014sideways}, the body moves perpendicular to how the body faces (Figure \ref{fig:HolonomicExpository}B). Thus, it is not essential that we walk in a manner that we always face the movement direction. So, we consider two ways of walking: walking while not always facing the movement direction (Figure \ref{fig:HolonomicExpository}C) and walking while always facing the movement direction (Figure \ref{fig:HolonomicExpository}D). We call these kinds of walking ``holonomic'' and ``non-holonomic'' respectively, borrowing this terminology from classical mechanics, control theory, and other prior work on locomotor paths \cite{arechavaleta2006nonholonomic,mombaur2010human,ne_mark2004dynamics}. The term `non-holonomic' simply implies that the system obeys a velocity constraint -- here, the constraint is that the body velocity direction is always along body orientation. Holonomic walking has no such velocity constraint and thus non-holonomic walking is a special case of holonomic walking.  We generalize the metabolic cost model of equation \ref{eq:MetRateFit}  to both these types of walking \cite{mombaur2010human}. Specifically, for holonomic walking, in which the body need not face the movement direction, we use the following revised metabolic cost of the form: 
\begin{equation} 
\dot{E} = \alpha_0 +   \alpha_1 v_b^2 + \alpha_2 \omega^2 +  \alpha_s v_s^2, \label{eq:MetRateFitHolonomic} 
\end{equation}
where $v_b$ is the body velocity component in the forward or anterior-posterior direction (in the direction that the body is facing) and $v_s$ is the body velocity component in the sideways or medio-lateral direction (perpendicular to how the body is facing); see Figure \ref{fig:HolonomicExpository}A. Here, we refer to equation \ref{eq:MetRateFitHolonomic} along with an additive cost for changing speeds \cite{seethapathi2015metabolic} as the `generalized metabolic cost model.' See Methods for further details. The rest of this article uses this generalized cost model to make predictions about how people walk in more complex situations involving turning.  Because holonomic walking is more general, we use this type of walking to make predictions in the rest of this main manuscript, but we also show results from non-holonomic walking in supplemental figures.

\begin{figure*}[htb!]
\begin{center}
	\includegraphics{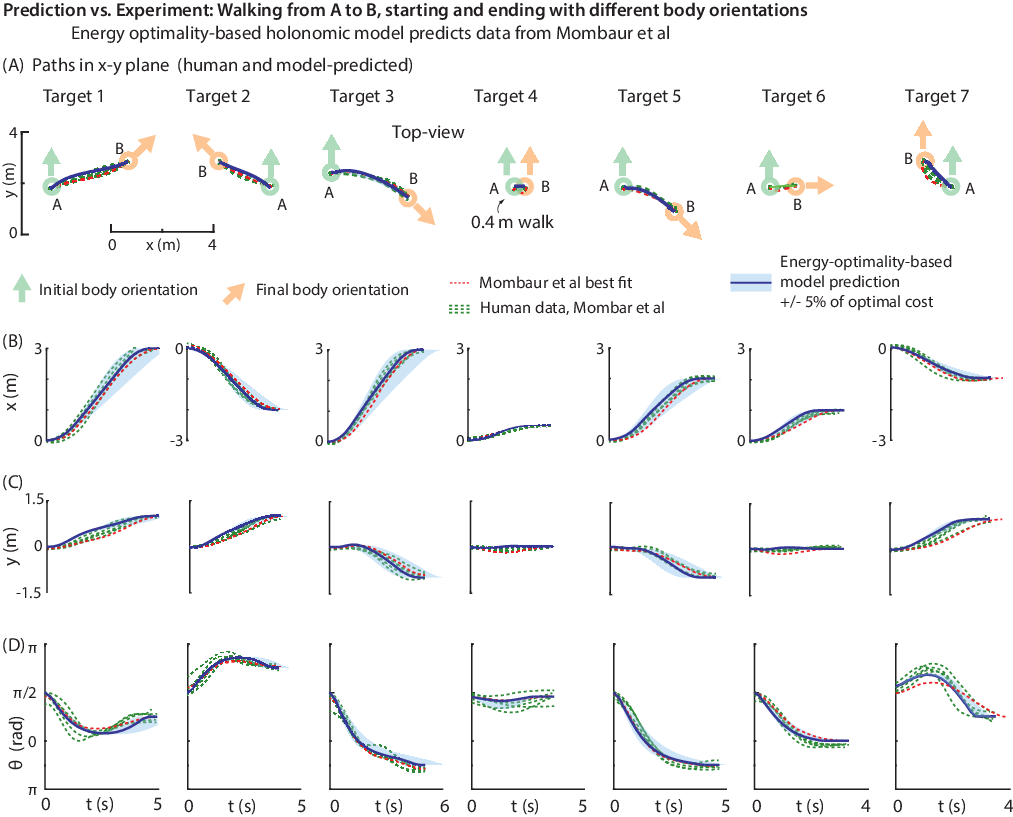}
\caption{\textbf{Prediction vs behavior: Path planning, starting and ending at rest.} (\textbf{A})  Mombaur et al \cite{mombaur2010human} asked subjects to walk short distances, starting at rest at point A and ending at rest at point B. The subjects had to start facing one direction (light green arrow) and end facing possibly another direction (orange arrow). (\textbf{B, C, D})  The body position $(x, y)$ and body orientation $\theta$ as a function of time. Holonomic model predictions are solid dark blue with a light blue band indicating trajectories withing 5\% of the optimum cost; experimental data are dashed dark green, and the best-fit model in Mombaur et al \cite{mombaur2010human}  is indicated in dashed red line. We see that our energy optimization-based model predictions mostly pass through the center of the experimental data. Just for targets 3 and 7, subjects started and ended with slightly different body orientations than prescribed, so these were used in the optimizations presented. See Supplementary Figures S1 and S2 for variants of this figure with alternate assumptions. Task 4 involves the  most striking difference between these holonomic model predictions versus the non-holonomic model predictions (Supplementary Figure S1); this task requires simply moving sideways by a short distance.}
\label{fig:MombaurTurningInitialFinalTangents}
\end{center}
\end{figure*}

\paragraph{Prediction vs. experiment: Navigating from A to B with constraints on initial and final direction.} Mombaur et al \cite{mombaur2010human} performed human subject trials in which the human started from rest at point A and ended at rest at point B, starting and ending with different body orientations (Figure \ref{fig:MombaurTurningInitialFinalTangents}A). The required body orientations were provided as arrows drawn on the ground. Subjects were not constrained in any other way, say by obstacles or time limits. Having different required body orientations at A and B requires the subjects to turn. For seven different end-point and body orientation combinations, we computed the metabolically optimal turning trajectory with our model and using trajectory optimization (see Methods and Supplementary Appendix). We performed two versions of the calculation, one holonomic and the other non-holonomic. Remarkably, we find that the holonomic model --- that is, allowing body orientation to be different from movement direction ---  predicts the time-progression of body position ($x$ vs $t$ and $y$ vs $t$) and body orientation ($\theta$ vs $t$) without fitting to this behavioral data (Figure \ref{fig:MombaurTurningInitialFinalTangents}B-D). In both the predicted optimal paths and the human paths, the body orientation is not always aligned with movement direction. Constraining body orientation to be aligned with movement (non-holonomic walking) produces worse predictions of  body position and orientation (Supplementary figure S3; see also \cite{mombaur2010human}).

\begin{figure*}[htb!]
\begin{center}
	\includegraphics{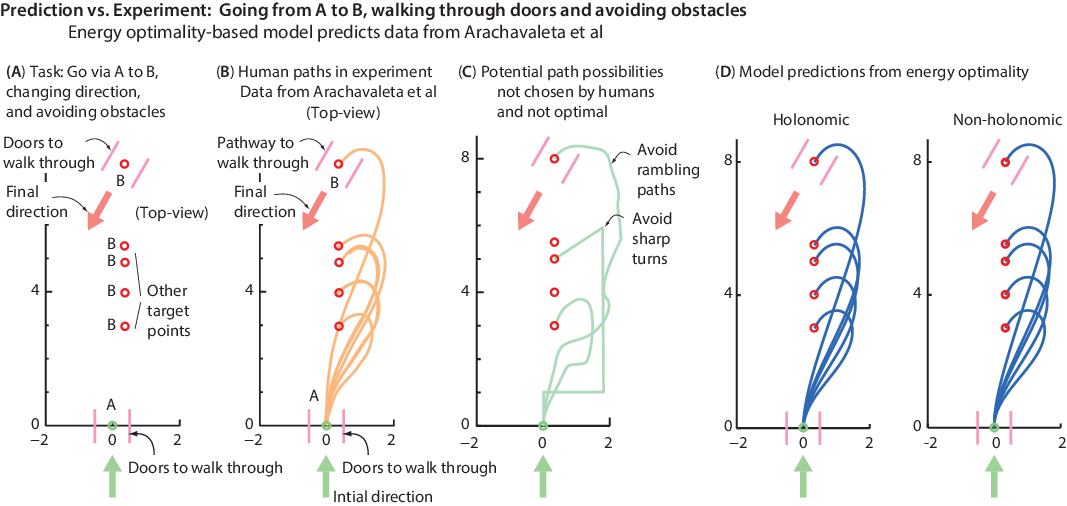}
\caption{\textbf{Prediction vs behavior: Navigating around and through doors.} (\textbf{A}) Humans were asked \cite{Are08} to walk through a doorway at point A  (pink parallel lines) to another doorway at point B, stopping 2 m beyond the second door. The second door B had five different locations as shown (red circles).  The walls of the doorways serve as obstacles to be avoided. (\textbf{B}) Human data redrawn from \cite{Are08}  are for head paths.  (\textbf{C})  A human or the model are capable of sharp turns and otherwise complex paths to achieve the task. c) But model predictions for the body path from energy optimality are qualitatively similar to human paths, despite not having to constrain the average velocity as in \cite{Are08}. See Supplementary Figure S5 for bands containing trajectories within 2\% of the optimal cost.}
\label{fig:TurningInitialFinalTangents}
\end{center}
\end{figure*}

\paragraph{Prediction vs. experiment: Navigation around and between two doorways.}
In another set of previous studies  \cite{arechavaleta2006nonholonomic,Hich07,Pham07}, researchers instructed human subjects to walk through two sets of doors A and B facing in different directions and separated by a few meters (Figure \ref{fig:TurningInitialFinalTangents}A). The subjects started from rest 2 m before A and ended 2 m beyond B. As in \cite{mombaur2010human}, humans chose smooth paths that gradually turn rather than, say, achieve the same task using sharp turns or too many direction changes (Figure \ref{fig:TurningInitialFinalTangents}B-C). Again, we used trajectory optimization to compute the energy-optimal way of performing this task. The resulting optimal trajectories are similar to the human trajectories in data (Figure \ref{fig:TurningInitialFinalTangents}), which are within 2\% of the optimal cost (Supplementary Figure S3). The predictions from the holonomic and non-holonomic models are almost the same, with the non-holonomic model taking a slightly wider turn near the door. For these longer distance bouts (compared to those in Figure \ref{fig:MombaurTurningInitialFinalTangents}), even when the walker is not constrained to be non-holonomic, it is energy optimal to be nearly non-holonomic -- that is, walk in a manner that the body nearly faces movement direction, as also observed in experiment \cite{arechavaleta2006nonholonomic}. Supplementary Figure S6 shows how the body movement direction $\beta$ closely follows body orientation $\theta$. The difference between the two angles ($\beta-\theta$) reduces with the distance traveled. 

An important constraint for the optimal path calculation here, in contrast to those for Figure \ref{fig:MombaurTurningInitialFinalTangents},  is that the body path does not intersect with the doors and has a minimum clearance from the doors. The clearance used is consistent with typical human dimensions \cite{Fruin1987} and also with  behavioral data \cite{dias2014pedestrian}. In the absence of such a clearance constraint, the optimal path ignores the walls of the doors and shows a sharper turn near the end-point B. Thus, explaining human behavior may require considering constraints such as avoiding obstacles (here, the doorways) in addition to minimizing energy-like cost functions.

\paragraph{Corollary: Shortest paths are not optimal and not used by humans.} In the previous two tasks (Figures \ref{fig:MombaurTurningInitialFinalTangents}, \ref{fig:TurningInitialFinalTangents}), our model predicted gradually turning paths as also exhibited by the human subjects. These paths are not the shortest paths between the origin and destination: the shortest paths for the tasks in Figure \ref{fig:MombaurTurningInitialFinalTangents} are straight lines whereas those for the tasks in Figure \ref{fig:TurningInitialFinalTangents} are straight line paths interspersed by a circular arc around a door. Thus, humans walk for a longer distance than necessary to save energy, even on flat terrain. This non-optimality of the shortest path is due to the additional cost for turning, without which, the shortest path would be energy optimal.

\begin{figure*}[ht!]
\begin{center}
	\includegraphics{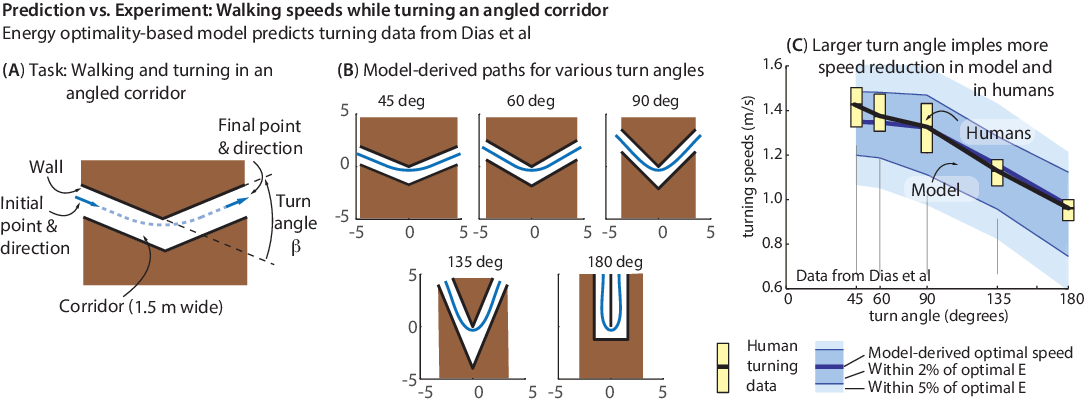}
\caption{\textbf{Prediction vs behavior: Navigating an angled corridor.} (\textbf{A})  A turning task, involving walking along a straight corridor and turning into another corridor angled with respect to the first. (\textbf{B})  Energy optimal paths for the task. Subject enters the first corridor and leaves the second corridor in a straight line path at their energy optimal speed and clears the wall. (\textbf{C})  Experimental data from Dias et al \cite{dias2014pedestrian} (mean $\pm$ s.d.) shows that the human speeds during the turn are lower for a larger turn angles, as also predicted by the model-derived energy optimal paths. The human speed distribution is entirely contained within the set of speeds within 2\% of the optimal cost.}
\label{fig:Dias}
\end{center}
\end{figure*}

\paragraph{Prediction vs. experiment: Humans slow down turning a corner while navigating an angled corridor.} A common everyday task  is turning a corner in an angled corridor (Figure \ref{fig:Dias}A). A previous study by Dias et al \cite{dias2014pedestrian} had subjects walk around angled corridors and measured the walking speeds during the turn. Again, using trajectory optimization, we computed the metabolically optimal path in such angled corridors, specifically computing the walking speed during the turn. We find that the optimal walking speed is lower during the turn and that this turning speed is lower for greater turn angles (Figure \ref{fig:Dias}B-C). Further, the experimentally observed human speeds from \cite{dias2014pedestrian} are almost identical to the model-predicted turning speed; the distribution of human turning speeds overlaps with the model-predicted band of speeds within 2\% of the optimal cost (Figure \ref{fig:Dias}C). Speed reductions during turns were similarly observed by Sreenivasa et al \cite{sreenivasa2008walking}, who considered turning  in a cyclical task that alternated between turning and straight line walking. In all these turning tasks, as predicted by the model, humans do not usually use `sharp turns' when smooth turns are possible. 

\paragraph{Rectilinear walking speeds are predicted as a special case.} The optimality criterion proposed here predicts rectilinear or straight-line walking phenomena as a special case. Specifically, it predicts that typical steady human walking speeds should be around $v_{\mathrm{opt}} = \sqrt{\alpha_0/ \alpha_1 } = 1.35$ m/s, which agrees with preferred walking speeds over longer distances in previous studies \cite{Sri09,bohannon2011normal,seethapathi2015metabolic}. Seethapathi and Srinivasan  \cite{seethapathi2015metabolic} used a straight-line walking metabolic model along with a cost for changing speeds to predict lowered walking speeds for short distance bouts. Our generalized metabolic cost model contains the model of \cite{seethapathi2015metabolic} as a special case (by construction) and thus also predicts that humans should walk systematically slower for shorter distances. Similarly, Handford and Srinivasan \cite{handford2014sideways} showed that when asked to walk sideways, humans walk close to the energy optimal sideways walking speed (about 0.6 m/s). Again, the generalized holonomic model contains the sideways walking model of  \cite{handford2014sideways} as a special case (by construction) and makes similarly low speed predictions for sideways walking. 

\paragraph{Prediction vs. experiment: To go sideways, walk sideways or turn and walk forward.} Humans do not usually walk sideways. Handford and Srinivasan \cite{handford2014sideways} showed that sideways walking costs three times more energy than walking forward at their respective optimal speeds. Now, consider a situation in which someone wants to go from A to B, but is initially (and finally) facing perpendicular to the line AB (Supplementary Figure S7). That is, they want to move sideways. e.g., they are working at a kitchen counter and want to move sideways. Should they walk sideways or turn by 90 degrees and walk facing forward? To make a prediction, we compare the cost of walking sideways and the cost of turning and walking forward and turning again. This comparison predicts that humans should walk sideways for distances less than $0.8$ m; for larger distances, walking sideways is more expensive than turning and walking facing forward. Indeed, target 4 in Figure \ref{fig:MombaurTurningInitialFinalTangents} asks subjects to travel sideways by 0.4 m, starting and ending facing forward \cite{mombaur2010human}. Subjects, as predicted, did not turn and walk forward, instead just stepped sideways while mostly facing forward.

\paragraph{The predictive power of minimizing metabolic energy versus other optimality hypotheses.} We have predicted a wide variety of experimentally measured human locomotor behavior, with no fitting to this behavioral data, by directly minimizing an empirically based metabolic model of walking. We now briefly compare the energy-cost-based hypothesis with some other hypotheses that have previously been explored. Smooth body trajectories can be predicted  by cost functions that maximize smoothness, such as acceleration, `jerk' (derivative of acceleration) or `snap' (second derivative of acceleration). Such cost functions have been successful in arm reaching \cite{Viv95}, but they have also been employed to predict walking trajectories \cite{Pham07,mombaur2010human}. However, such smoothness maximizing cost functions, taken seriously, produces two un-ecological predictions. First, they cannot predict the velocity at which people move: specifically, the optimal way to minimize jerk or snap is to perform a task with infinitesimal speed over an arbitrarily long period of time. See Supplementary Appendix (section S4) for a mathematical account of using these smoothness objectives. Thus, such smoothness maximizing cost functions require a constraint on the average velocity of the task to produce meaningful results. In contrast, our metabolic energy approach naturally produces an optimal velocity for any task, without having to constrain it. A second consequence of the jerk minimization hypothesis, even with a velocity constraint, is that it produces `scale-invariant solutions'. That is, for the tasks in Figure \ref{fig:TurningInitialFinalTangents}A, it produces paths that `look the same' for a 10 meter walk versus a 10 km walk, producing many-km long path excursions that humans would never use. 

Arachavaleta et al \cite{Are08} used an objective function equal to the integral of $(v^2 + \dot{\kappa}^2)$ over the path, where $\dot{\kappa}$ is the rate of change of path curvature. Again, minimizing this objective without a further constraint predicts that humans should move at infinitesimal velocity. So in their calculations, Arachavaleta et al \cite{Are08} constrained the total time. Thus, these minimization hypotheses require further assumptions about human behavior, which need not be made with the more parsimonious energetics-based approach. 

Finally, Mombaur et al \cite{mombaur2010human} used a model-fitting procedure to select an objective function that best predicts human walking trajectories in their experiments: their objective function terms related to linear and angular velocity acceleration, work, jerk, and task time duration. However, their best-fit cost function, because it is dominated by linear and angular acceleration terms, cannot explain observed speeds for walking in circles, turning in place, or walking in a straight line for longer distances.  Thus, each of these previous cost functions could be considered as overfit to one or two experimental conditions and cannot predict all the diverse phenomena predicted by our model. On the other hand, the true objective may well contain terms analogous to those used in these prior work \cite{Pham07,mombaur2010human,Are08} in addition to those in our metabolic model; for instance, smoothness-promoting terms related to acceleration or jerk, as a proxy for costs related to muscle forces and force rates \cite{wong2020energetic,kram2000muscular,Dok05}. Such terms may be useful in explaining smoother speed changes (than our model) when speed changes are needed, for instance, during gait initiation and termination \cite{miff2005temporal}.

\section*{Discussion}
We have provided a unified theoretical account of human navigation paths and speeds. We first measured the energetics of humans walking with turning and showed that the energy cost has a substantial dependence on path curvature. Turning increases the locomotor cost and this turning cost increases with the turning speed. We then used the experimentally-derived metabolic energy model to predict energy optimal walking behavior in various contexts, explaining many disparate experimentally-measured and ecological human locomotor behavior, some via new experiments and some by comparison with data from prior experimental studies on human navigation. Importantly, our theoretical account also makes predictions about straight-line locomotion as special cases, predicting steady walking speeds and short distance speeds when turning is not required. Thus, we have presented a unified theoretical account of both straight line and more general curvilinear human walking. 

We showed that the tendency to slow down when turning and slow down more when turning in a smaller radius can be explained by energy optimality. One alternate hypothesis for slowing down while turning is to avoid slipping. While fast-moving cars and bicycles slow down on curves to avoid slipping, humans in our experiments were far from any danger of slipping. We estimated the foot-ground friction coefficient as $\mu = 0.6 $ to $1.2$, giving friction cone angles of 30 to 50 degrees ($=\tan^{-1}\mu$). But the maximum leg angle in our circle walking (1 m at 1.5 m/s) was 12 degrees, much less than the friction cone angles, giving a large safety factor from slipping. 

The speed reductions found here due to path curvature are reminiscent of speed reductions when humans try to run as fast as possible around a circle \cite{greene1985running,chang2007limitations}. However, these tasks are different in one important respect; while our subjects are able to walk faster or slower compared to their preferred speeds, maximal speed running does not have this property (by definition). Nevertheless, there may be mechanistic similarities between the two cases. As was initially thought for maximal running \cite{greene1985running,chang2007limitations}, we might hypothesize that the increased walking metabolic cost may be due to the necessity for producing a centripetal force; see \cite{kram2000muscular,Grif03,roberts1998energetics,Dok05,hubel2015children,biewener2004muscle} for accounts of force costs in locomotion. However, using elementary formulas for centripetal accelerations \cite{greene1985running}, we estimated that for the fastest walking speed at the smallest radius, the leg force requirements increased only by 1.3\% (see Supplementary Information Appendix, section S3). This small force increase cannot directly account for the nearly 50\% increases in metabolic costs we have measured. Similarly, if we conceptualize the additional centripetal acceleration as increasing the `effective gravity' (resulting in the increased leg forces) by 1-2\%, then the leg work requirements would also increase by a similar small percentage and not by over 50\% \cite{Sri11,donelan1997effect,grabowski2005independent}. One might speculate that the larger increases in metabolic rate may be partly due to recruitment of additional muscles that may not be at their optimal operating regimes \cite{chang2007limitations}. 

The cost of turning could also be due to the work necessary to turn the body orientation, given that the body has a non-zero rotational inertia \cite{qiao2014compensations}. Per our assumed metabolic cost model form, the metabolic cost of turning in place was $\dot{E} = \alpha_0 + \alpha_2 \omega^2$. Turning in place only requires body orientation changes and does not require centripetal forces. Thus, this cost model for turning in place is consistent with the premise that most of the increased cost due to turning (namely, $ \alpha_2 \omega^2$) is to accomplish body orientation changes. See  Supplementary Information Appendix, section S3. The increased metabolic cost could potentially be understood by measuring the body movements and the ground reaction forces \cite{orendurff2006kinematics,Court03b,turcato2015generation,orendurff2006kinematics}, performing inverse dynamics \cite{falisse2019rapid}, and examining which joints have greater torques or perform more work.

Our metabolic cost model is empirical in nature, so that to generalize it to different situations (e.g., running or uneven terrain), we may need to repeat the metabolic experiments for that new situation. An alternate path to potential generalization is via a 3D dynamical model of a biped \cite{Miller2014ComparisonWalking,handford2016robotic}, first showing that it is able to predict the walking metabolic measurements here (and along the way, also incorporating the stepping dynamics, ignored here for parsimony). But, such 3D biped models may still not generalize to other human populations, for instance, those with movement disorders, whether due to musculoskeletal or neurological differences, because predictive understanding of movement behavior with movement disorders without task- or population-specific fitting is largely open. On a related note, we have assumed left-right symmetry in fitting the metabolic cost model (equation \ref{eq:MetRateFit}), but bodily asymmetries may translate to energetic asymmetries in turning eg. in unilateral amputee populations. Accounting for such asymmetries in the metabolic model, one can then examine whether no turning ($\omega = 0$) still minimizes the metabolic rate for speed $v$ and whether a straight line path remains energy optimal.

One could conjecture that there are greater stability issues while turning and that these stability issues contribute to the subjects being cautious and lowering their speed. However, such conjecture seems unnecessary as energy minimization seems to largely explain the speed reductions. An open question in movement control is to what extent humans prioritize energy or effort on the one hand and stability or robustness to uncertainty on the other \cite{saglam2014quantifying}. This question is beyond the scope of this study, as we have based our behavioral predictions on empirically derived energy costs. The experimentally measured energy costs already include any trade-offs that humans had to make to walk efficiently and stably \cite{seethapathi2019step,joshi2019controller}. Thus, we should distinguish our empirical optimality criterion from the theoretical limit of true energy optimality, requiring perfect control and the absence of perturbations or uncertainty.

In the real world, there may be other additional concerns that may make a human walk faster than energy optimal, for instance, a cost for time or constraints on time taken to complete the movement \cite{LongSrinivasan2013,summerside2018contributions}. Here, we have considered conditions where no such explicit time-pressure exists. Deviations from deterministic energy optimality may also be due to  optimality in a Bayesian or probabilistic sense \cite{kording2004bayesian} in the presence of uncertainly or due to interactions with or navigating around another moving human.

It is sometimes argued that energy optimality or energy economy may be useful only in `steady state tasks' as opposed to `transient tasks' or may only be useful when the energy saved is substantial \cite{seethapathi2015metabolic}. Here, we have shown broad agreement of behavior with empirical energy optimality in short transient tasks (e.g., the turning part a corner) that consume a very small amount of energy. For instance, we estimate the total energy cost of a turn in the angled corridor (Figure \ref{fig:Dias}A-C) to be about 10 J/kg. So, the savings relative to a non-optimal turn (e.g., not slowing down or using a sharper turn) are a small fraction of this cost, equivalent to just over a second of resting energy expenditure. Here, as in many of the situations we considered, the experimentally observed behavior were within 2\% of the optimal energy costs from the model, corresponding to similarly small amounts of energy differences (0.2 J/kg, equivalent to resting energy expenditure for one seventh of a second). That energy optimality provides an account of diverse transient behavior with low energy requirements is perhaps an indication of how much the nervous system values energy savings, all else being equal. Our results are agnostic to how the near-energy optimality is achieved, whether it is hard-wired evolutionarily, acquired while learning to walk during childhood, or if the energy optimal trajectories are computed in real time by the nervous system \cite{selinger2015humans,wong2017contribution}.  It is likely a mixture of all such mechanisms: energy optimality correctly predicted lowered walking speeds for shorter distances (a task with low total energy \cite{seethapathi2015metabolic}), walking speeds in sideways walking (an uncommon task \cite{handford2014sideways}), and stride frequency in the presence of a external exoskeleton (an uncommon task with dynamic changes in the energy landscape \cite{selinger2015humans,selinger2019humans}). 

The neural mechanisms underlying such energy optimality or locomotor navigation are yet to be elucidated  \cite{edvardsen2020navigating}. Further, the aspect of navigation we focused on were path and speed selection when all information about the world is fully available, and not aspects involving sensing, information gathering, and course correction, or topological aspects in which one among multiple (potentially locally optimal) paths around obstacles need to be selected \cite{warren2008behavioral,baxter2020route}. 

Our results suggest further behavioral experiments to test model predictions and inform improvements to the model: for instance, walking through via points with freedom to choose the intervening path, comparing walking sideways versus turning, walking around obstacles, walking around moving obstacles (such as other people), etc. We obtained a cost of spinning in place by extrapolation, whose accuracy can be improved by using smaller radii. Because of this extrapolation, we would a priori expect the current model to be less reliable in tasks that require sharp direction changes. Mechanisms for the turning cost could be probed by predicting and testing experimentally how the cost varies after various manipulations of the system: adding mass or moment of inertia to the trunk or the legs \cite{qiao2014compensations} or providing centripetal forces via a tether as in \cite{chang2007limitations}. 
Our empirical metabolic cost model could be tested further by energetic measurements of humans walking on sinusoidal paths, achieved by moving side to side on treadmills. Such curvilinear paths with non-constant curvature may allow us to disambiguate energy cost dependence on $v$ and $\omega$ versus their derivatives, which may be more significant for more unsteady tasks. 

The metabolic model for curvilinear locomotion presented here may help improve the estimates of daily metabolic expenditure, say, using wearable devices. Studies have found about 20\% of steps in household settings \cite{Sedgeman1994} and 35-45\% of steps in common walking tasks in home and office environments  involve turns \cite{glaister2007video}, so neglecting the cost of turning may result in erroneous estimates of energy expenditure. Further ambulatory studies tracking people over many days  with wearable sensors \cite{daley2016preferred} could help estimate the magnitude of this error.  

As noted earlier, reduced maximal running speeds around a circle have been previously measured \cite{greene1985running,chang2007limitations}, but sub-maximal running studies measuring metabolic costs have not been performed. Generalizing our metabolic model to running may better help estimate the metabolic cost during sports (e.g., soccer), which involve extensive speed and direction changes. Understanding human locomotion while turning would also be a useful tool in rehabilitation or assistive robotics.  Robotic legs, prostheses, and assistive devices are often designed with an emphasis on straight line walking. So, better understanding turning mechanics (for instance, by providing targets of turning performance) may inform designing for real world scenarios where curvilinear locomotion is essential.

In conclusion, through experiments and mathematical models, we have provided a unified theoretical predictive account of human walking in non-straight-line paths and with turning from the perspective of energy optimality. We have suggested further experiments to test model predictions, to inform model improvements, and to generalize the models to other populations and other tasks. 

\matmethods{

We performed three different experimental studies, one for measuring the metabolic cost of turning and two for measuring human behavior while turning.   We performed multiple model-based optimization calculations to predict energy optimal trajectories and speeds under different task constraints to compare with a number of different behavioral experiments, including our own. 

\paragraph{Experiment: Metabolic cost of humans walking in circles.}  All experiments were approved by the Ohio State University's institutional Review Board and all subjects participated with informed consent. Subjects were instructed to walk along circles drawn on the ground (Figure \ref{fig:MetExpt}A). The subjects were instructed to keep the circle directly beneath their feet or between their two feet, but not entirely to one side of their feet. All subjects walked with the circle between their feet with non-zero step width. We used four different circle radii ($R = 1, 2, 3, 4$ m, $N_{\mathrm{radii}} = 4$). At each radius, subjects performed four walking trials, each with a different constant tangential speed $v$ in the range 0.8-1.58 m/s, resulting in $N_\mathrm{trials} = 16$ trials per subject; one subject performed fewer trials ($N_\mathrm{trials} = 13$). Tangential speeds were prescribed by specifying a duration for each lap around the circle. A timer provided auditory feedback at the end of every half lap duration (for $R = 3, 4$ m) or full lap duration (for $R =1, 2$ m), so that subjects could speed up or slow down as necessary. Within a few laps of such auditory-feedback-driven training, subjects walked at the desired average speed, completing each lap almost coincident with the desired lap time. Subjects maintained the speed with continued auditory feedback for 6-7 minutes: 4 minutes for achieving biomechanical and metabolic steady state and 2-3 minutes for computing an average metabolic rate $\dot{E}$.  Subjects used clockwise or counter-clockwise circles as preferred. Subjects were instructed to walk and never jog or run; all subjects always walked. Overall, this resulted in nearly 35 hours of subjects walking in circles.

The trial order was randomized over speed and radius for seven subjects (mass $77.3 \pm 10$ kg, height $1.79 \pm 0.05$ m, mean $\pm$ s.d. and age range 22-27), whose analyses are presented in detail in the Results section. For ten other subjects (mass $73 \pm 14$ kg, height $1.75 \pm 0.13$, mean $\pm$ s.d. and age range 21-27), the trial order increased monotonically in speed and radius. Nevertheless, the overall regression relations were similar when both sets of data were analyzed in the same manner, suggesting no large order effect. Metabolic rate per unit mass $\dot{E}$ was estimated during resting and circular walking using respiratory gas analysis (Oxycon Mobile with wind shield, $<$ 1 kg): $\dot{E} = 16.58 \, \dot{\mathrm{V}}_{\mathrm{O}_2} + 4.51 \dot{\mathrm{V}}_{\mathrm{CO}_2}\, \mathrm{W/kg}$ with volume rates $\dot{\mathrm{V}}$ in ml.s$^{-1}$kg$^{-1}$  \cite{Brock87}. 
Subjects exhibited small but systematic differences between prescribed lap times and lap times estimated from video. So, in subsequent analyses, we used corrected values for $v$ and $\omega$ by using the estimated lap times in their calculation (speed = circumference/lap time, angular speed = 2$\pi$/lap time). See \textit{Supplementary Material} for the collected data, including these corrections. To improve estimates of the steady state metabolic rate and to test if the transients have subsided, we fit an exponential to the last 3 minutes of metabolic data and determined the extrapolated steady state \cite{selinger2014estimating,zhang2017human}. This extrapolation resulted in less than 2\% changes in any of the coefficients in equation \ref{eq:MetRateFit}, compared to the standard procedure of using the mean metabolic rate over the last 2 or 3 minutes. This comparison suggests that the standard procedure suffices. Reported significance of metabolic differences across radii were tested via one-sided t-tests with a Bonferroni correction for multiple comparisons. Linear regressions used \texttt{fitlm} in MATLAB.

\paragraph{Experiment: Preferred walking speeds in circles.} Subjects' preferred walking speeds were measured by asking them to walk in a straight-line and along circles \cite{orendurff2006kinematics} of radii ($R$) equal to 1\,m, 2\,m, 3\,m, and 4\,m at whatever speed they found comfortable (Figure \ref{fig:MetCostPerDistannceEtc}A); the subjects walked for about 100\,m in each of these trials (four 4\,m laps, eight 2\,m laps, etc.). The subjects were told beforehand how many laps they would need to complete and were asked to do whatever felt normal or natural. We measured the time duration $T$ for the second half of the walk from video and estimated the average tangential speed as the distance traveled around the circle divided by the time duration; that is, the average tangential speed is $(2\pi R \cdot n_\mathrm{laps})/T$, where $n_\mathrm{laps}$ is the number of laps considered. Two trials were performed for each radius and all trials were in random order of radii. The subject population for these trials  (mass 73.6 $\pm$ 10 kg, height 1.74 $\pm$ 0.13 m, age 22.6 $\pm$ 1.7 years, 90 trials with $N = 9$) was distinct from those in the previous experiment. 

\paragraph{Experiment: Preferred turning-in-place speeds.} Subjects' (mass 73.4 $\pm$ 11 kg, height 1.75 $\pm$ 0.10 m, age 26 $\pm$ 5 years, $N = 10$ distinct from earlier samples) preferred turning speeds were measured by asking them to turn in place by 90, 180, 270, and 360 degrees and do whatever feels natural (Figure \ref{fig:TurningInPlace}A). Three trials were performed for each turn angle and the trial orders were randomized. Subjects were free to turn clockwise or counter-clockwise in any trial. The average angular velocity of turn was computed by estimating the time taken for turning from video and using the prescribed turn angle. Our goal was to compare preferred turning speeds with model-based predictions of steady state turning speeds, that is, speeds not affected by the starting and stopping. We found that the average speeds for 270 and 360 degree turns were similar  (Figure \ref{fig:TurningInPlace}), so we infer that these speeds are close to the steady state speeds. Thus, we use the turning speeds for 270 and 360 degree turns to test predictions of steady state turning speeds. 

\paragraph{Model: Metabolic cost of arbitrary walking paths.} Our walking-in-circles metabolic experiments constrained the circular paths that the feet travel on rather than the paths that the body travels in. Thus the radius $R$ and the velocity $v$ in equation \ref{fig:MetExpt} correspond to the mean trajectory of the feet rather than the center of mass. If the foot travels in a circle of radius $R$ and has an effective tangential velocity $v$, the body center of mass travels in a circle of smaller radius $R_b$ and slightly lower tangential velocity $v_b$. This is because the body leans into the circle, so that the hip is closer to the center of the circle than the foot \cite{greene1985running}. We first obtain a body-based description of the empirical metabolic cost: $\dot{E} =  \alpha_0^\prime + \alpha_1^\prime  v_b^2 + \alpha_2^\prime \omega_b^2$, where $\omega_b = \omega = v/R = v_b/R_b$ for circle walking. We find $\alpha_0^\prime = 2.32$ W/kg,  $\alpha_1^\prime = 1.28$ W/kg/$($ms$^{-1})^2$, and  $\alpha_2^\prime = 1.02$ W/kg/$($rad.s$^{-1})^2$. These coefficients explain the metabolic data roughly as well as the original model (explaining about 88\% variance).

Any non-circular or non-straight-line walking path can be described as a curve with constantly changing curvature. That is, each point on the curve has a distinct radius of curvature. If we assume non-holonomic walking, in which the body always faces the movement direction, we can directly apply a metabolic rate of the form $\dot{E} = \alpha_0^\prime + \alpha_1^\prime v^2 + \alpha_2^\prime \omega_b^2$, where $v_b$ is the instantaneous body velocity,  $\omega_b = v_b/R_b$ is the body angular velocity, and $R_b$ is the instantaneous radius of curvature. To generalize to holonomic walking, that is, allowing the body to not always face the velocity direction, we distinguish between the body velocity component along the body orientation $v_f$ (forward) and the body velocity component perpendicular to the body orientation $v_s$ (sideways). We then use a metabolic cost model of the form:  $\dot{E} = \alpha_0^\prime + \alpha_1^\prime v_f^2  + \alpha_2^\prime \omega_b^2 + \alpha_s^\prime v_s^2 $. This is the same as equation \ref{eq:MetRateFitHolonomic}. Here, the new term $\alpha_s^\prime v_s^2$ is the incremental cost of sideways walking  \cite{handford2014sideways} . Handford and Srinivasan  \cite{handford2014sideways} estimated the coefficient $\alpha_s^\prime$ roughly 7.8 W/kg/(ms$^{-1})^2$.  

Because walking along arbitrary paths may involve or require changing walking speeds, we include the additive metabolic cost of changing speeds, previously characterized by Seethapathi and Srinivasan \cite{seethapathi2015metabolic}. This study showed that accounting for this cost predicts lower speeds for short distance walking bouts using energy optimality, as in humans \cite{seethapathi2015metabolic,miff2005temporal}. Equation \ref{eq:MetRateFitHolonomic} in addition to this work-based cost for changing speeds is what we refer to as the `generalized metabolic cost model' in this manuscript. See Supplementary Appendix for more details about the metabolic cost model.

\paragraph{Model: energy-optimality-based behavioral predictions.} We compare measured experimental human behavior in a number of different walking tasks to the energy optimal walking behavior predictions. The energy optimal walking behavior is obtained by minimizing the total metabolic cost of the walking task. For the simplest two tasks, namely, walking in circle and turning in place (Figures \ref{fig:MetCostPerDistannceEtc}-\ref{fig:TurningInPlace}), the optimization assumes steady state and requires only basic calculus. So the complete analytical reasoning for the prediction is provided in the Results section. For more complex walking tasks where the walking path is not pre-determined (Figures \ref{fig:MombaurTurningInitialFinalTangents},\ref{fig:TurningInitialFinalTangents},\ref{fig:Dias}), we solve for the total time duration, body position, body orientation, and their derivatives as functions of time using numerical trajectory optimization methods  \cite{Sri11}. For this trajectory optimization, we use the metabolic cost function described in the previous paragraph. We perform two versions of the optimization, holonomic and non-holonomic, with the latter obeying the constraint that the body always faces the velocity direction. We solve additional optimization problems to obtain trajectories within a certain percent of the optimal cost. See  \textit{Supplementary Information} for more mathematical details of the numerical optimization. 
}

\showmatmethods{} 

\acknow{This work was supported by NSF grant CMMI-1254842 and its writing in part by the NIH grant R01GM135923-01.  We thank Carmen Swain and Blake Holderman for help regarding metabolic equipment during early pilot testing, Alison Sheets for comments on an early draft, and V. Joshi for help with some experimental setups. }

\showacknow{} 

\bibliography{WalkWithTurning}

\end{document}